\documentclass[superscriptaddress,twocolumn,showpacs,preprintnumbers,amsmath,amssymb]{revtex4}
\usepackage{graphicx}
\usepackage{dcolumn}
\usepackage{bm}
\usepackage{color}
\usepackage{ulem}
\usepackage{epsfig}

\begin{document}
\title{Broadband multimode fiber spectrometer}
\author{Seng Fatt Liew}
\author{Brandon Redding}
\affiliation{Department of Applied Physics, Yale University, New Haven CT 06520, USA.}
\author{Michael A. Choma}
\affiliation{Deparment of Radiology and Biomedical Imaging, Yale University, New Haven, CT 06520.}
\affiliation{Department of Pediatrics, Yale University, New Haven, CT 06520.}
\affiliation{Department of Biomedical Engineering, Yale University, New Haven, CT 06520.}
\affiliation{Department of Applied Physics, Yale University, New Haven CT 06520, USA.}
\author{Hemant D. Tagare}
\affiliation{Deparment of Radiology and Imaging Science, Yale University, New Haven, CT 06520}
\affiliation{Department of Biomedical Engineering, Yale University, New Haven, CT 06520.}
\author{Hui Cao}
\affiliation{Applied Physics Department, Yale University, New Haven CT 06520, USA.}







\begin{abstract}
A general-purpose all-fiber spectrometer is demonstrated to overcome the trade-off between spectral resolution and bandwidth. 
By integrating a wavelength division multiplexer with five multimode optical fibers, we have achieved 100 nm bandwidth with 0.03 nm resolution at wavelength 1500 nm. 
An efficient algorithm is developed to reconstruct the spectrum from the speckle pattern produced by interference of guided modes in the multimode fibers. 
Such algorithm enables a rapid, accurate reconstruction of both sparse and dense spectra in the presence of noise. 
\end{abstract}


\maketitle



Spectrometers have been widely used for spectroscopy-based sensing and imaging applications. 
In addition to the conventional one-to-one spectral to spatial mapping in grating or prism spectrometers, complex spectral to spatial mapping has been explored to make compact spectrometers with fine resolution \cite{Xu_OE, Dogariu_OL, Hang_AO, Brandon_NP, Mazilu_OL14, Menon_OE, Hanson_OL15, Wan_NatCom15,Yang_OL15, Bao_Nat}.
We recently showed that a multimode fiber (MMF) can function as a general purpose spectrometer with ultrahigh resolution \cite{Brandon_OL,Brandon_OE,Brandon_AO,Brandon_Optica}. 
The interference of guided modes in a MMF forms a seemingly random speckle pattern at the output, which is distinct for light at different wavelength, thus providing a spectral fingerprint.
After calibrating the wavelength-dependent speckle patterns, an arbitrary input spectrum can be reconstructed from the output intensity profile \cite{Brandon_OL,Brandon_OE}.
The spectral resolution is determined by the sensitivity of speckle pattern to a change in wavelength, which scales inversely with the fiber length. 
A 100 meter long MMF can provide the resolving power that exceeds the state-of-the-art grating spectrometer \cite{Brandon_Optica}. 
Since the MMF can be coiled to a small volume and has little loss, the fiber spectrometer offers the advantages of high sensitivity, compact size, light weight, and low cost.  

\indent However, the MMF spectrometer suffers a trade-off between spectral resolution and bandwidth, similar to a conventional spectrometer. 
The bandwidth of a MMF spectrometer is determined by the number of independent spectral channels that can be measured simultaneously, which is usually limited by the total number of guided modes in the MMF. 
In principle, one could increase the bandwidth by using a larger core fiber which supports more spatial modes; however, when the fiber spectrometer is used to measure spectrally dense signals, the speckle contrast is reduced to $\sim N^{-1/2}$, where $N$ is the number of uncorrelated speckle patterns produced by signals from independent spectral channels. 
An accurate recovery of the spectrum requires the speckle contrast to be greater than the measurement noise, which limits the spectral bandwidth.  

The resolution-bandwidth trade-off poses a serious issue for many practical applications. 
For example, in the spectral-domain optical coherence tomography (OCT), the spectral resolution of the spectrometer determines the imaging depth while the spectrometer bandwidth sets the axial spatial resolution \cite{OCT_RPP}. 
The trade-off between spectral resolution and bandwidth therefore translates directly to a trade-off between imaging resolution and spatial range. 
In order to replace the bulky grating spectrometer currently used in the spectral-domain OCT systems with an all-fiber spectrometer, a broadband operation must be achieved without sacrificing spectral resolution.  
 
In this Letter, we overcome the resolution-bandwidth trade-off by taking advantage of wavelength-division multiplexers (WDM) that have been previously developed for fiber communications. 
A commercial WDM separates a broadband signal collected by a single mode fiber (SMF) into multiple SMFs, each receiving a subset of the input spectrum. 
Since the WDM is already fiber-based, it can be easily integrated with the MMF spectrometer to provide very broadband operation with fine spectral resolution.
In particular, we demonstrate a single-shot measurement of dense spectra with 100 nm bandwidth and 0.03 nm resolution at the wavelength of 1500 nm. 
Such performance meets the typical requirement for spectral-domain OCT applications, and our all-fiber system is much more compact and portable compared to the commonly used grating spectrometer. 
In addition, we have developed an efficient spectral reconstruction algorithm that enables a rapid, accurate spectral reconstruction against measurement noise. 
Such algorithm is applied successfully to recover different types of spectra, such as discrete lines, continuous bands, or a combination of these features over a wide spectral range. 
This broadband, high-resolution, all-fiber spectrometer may enable a hand-held OCT imaging system, thereby improving the accessibility of this emerging imaging modality. 


\begin{figure}[htbp]
	\centering
	\includegraphics[width=\linewidth]{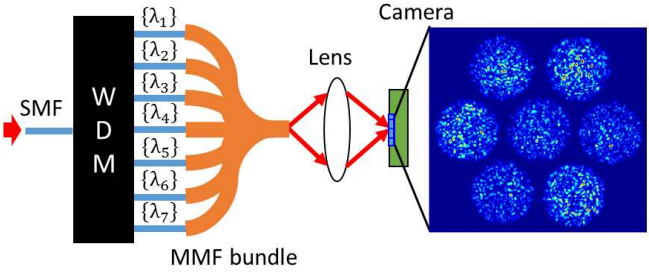}
	\caption{(Color online) A wavelength division multiplexer (WDM) and 1-to-7 fan-out fiber bundle are used to increase the spectral bandwidth of a fiber spectrometer. Each of the output single mode fiber of the WDM is connected to a multimode fiber in the bundle. The common end fiber bundle facet is imaged onto a InGaAs camera to record the speckle patterns.  }
	\label{fig1}
\end{figure}

\indent We use a commercial WDM module (MPS-2800) with one input SMF and eight output SMFs. 
The input light in the wavelength range of 1460nm -- 1620nm is split to 8 bands, each covering about 20 nm width. 
Seven output SMFs are connected to a bundle of seven MMFs [Fig. \ref{fig1}]. 
Each MMF is 2 meter long, the core diameter is 105 $\mu m$, and the numerical aperture is 0.22. 
There are about 1000 guided modes at the wavelength $\lambda$ = 1500 nm. 
The seven MMFs are closely packed at the common end of the 1-to-7 fan-out fiber bundle, the speckle patterns from all of them are captured simultaneously by a monochrome InGaAs camera (Xenics Xeva 1.7-640). 
A linear polarizer is placed in front of the camera to measure the polarized speckle with higher contrast.    

\indent To characterize the spectral resolution of the MMF, we coupled a tunable laser (Agilent 81600B) to the SMF input port of the WDM and measure the spectral correlation of the output speckle pattern with wavelength tuning. 
The half-width-at-half-maximum of the correlation function is $\delta\lambda$ = 0.03 nm, meaning the speckle correlation is reduced to $50\%$ with a wavelength change of 0.03 nm.   
This value provides an estimate for the spectral resolution, as a noticeable difference in the speckle pattern allows us to distinguish two wavelengths. 
The speckle pattern from a single MMF contains approximately 700 speckles, which gives the number of independent spectral channels $M$ that can be measured in parallel. 
Thus the spectral bandwidth of a single MMF is $M\times\delta\lambda$ = 21 nm. 
Previously, in order to increase the spectral bandwidth, we had to reduce the spectral resolution by using a shorter fiber \cite{Brandon_OE, Brandon_Optica}. 
Here, we keep the resolution and increase the bandwidth by using a WDM and multiple MMFs. 
The large number of pixels, afforded by modern 2D cameras, allows the speckle patterns from all the cores to be recorded in a single shot [Fig. \ref{fig1}].

Before using the MMF spectrometer, we must calibrate the spectral-spatial mapping. 
The transfer matrix, $T$, connects the input spectrum to the output speckle pattern, namely, $I = T \cdot S$, where $I$ is a vector describing the discretized speckle pattern and $S$ the discretized spectrum.
For calibration, we scan the wavelength of the tunable laser in the step of 0.02 nm and record the output speckle pattern at each wavelength which forms one column in $T$.  
Due to the limited tuning range of the laser, we use only five spectral bands of the WDM which covers the wavelength from 1520 nm to 1620 nm. 
The transfer matrix contains 5000 spectral channels and 11000 spatial channels in total.  
To reduce the measurement noise, we repeat the calibration procedure and average the transfer matrix over ten sequential measurements.

\begin{figure}[htbp]
	\centering
	\includegraphics[width=\linewidth]{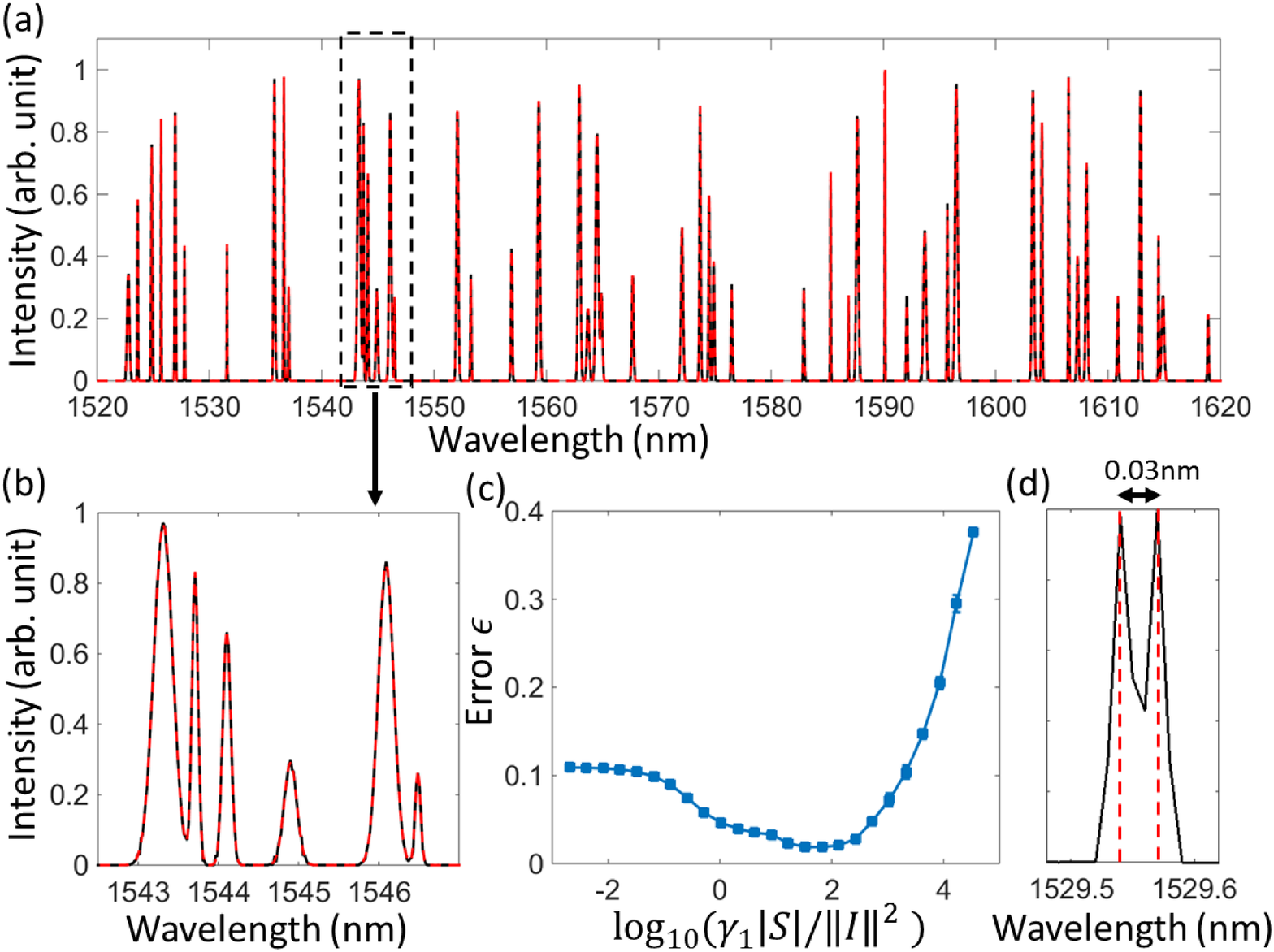}
	\caption{(Color online) (a) A reconstructed spectrum that contains 50 Gaussian-shape peaks with varying width and amplitude over 100 nm range. The black-solid line is the reconstructed spectrum $S'$ and the red-dashed line is the input spectrum $S$. The reconstruction error is $\epsilon = 0.03$. (b) Expansion of part of the spectrum in (a) shows the spectral shape of each peak in the reconstructed spectrum matches very well with the input spectrum. (c) Reconstruction error $\epsilon$ as a function of $\gamma_1$. The minimum of $\epsilon$ gives the optimum value of $\gamma_1$ for spectra reconstruction. (d) Two lines separated by 0.03 nm are resolved.  }
	\label{fig2}
\end{figure}

\indent After calibration, we test the MMF spectrometer by reconstructing the spectrum of a probe signal from its speckle pattern. 
To characterize the spectrometer's performance in reconstructing different types of spectra, we synthesize the speckle pattern of the probe signal using another transfer matrix $T'$ which was measured separately from the one for calibration, and $I = T' \cdot S$.
If there were no measurement noise, the spectrum could be easily retrieved from the inverse of the transfer matrix.  
However, in the presence of measurement noise, the matrix inversion is ill-conditioned, causing error in spectrum reconstruction.
Previously truncated inversion and nonlinear optimization methods have been adopted to reduce the reconstruction error \cite{Brandon_OE,Brandon_AO}. 
Here, we develop a more efficient and accurate algorithm that is robust against measurement noise.

\indent We start with recovering spectrally sparse, narrow-band signals. 
Instead of minimizing the norm $||I-T \cdot S' ||_2^2  \equiv \sum_j |I_j - \sum_i T_{j \, i} S'_i|^2$, we minimize $||I-T \cdot S' ||_2^2 + \gamma_1 ||S'||_1 $, where $S'$ is the reconstructed spectrum, $||S'||_1 \equiv \sum_i |S'_i|$, and $\gamma_1$ is a parameter. 
The additional term $\gamma_1 ||S'||_1$ represents a penalty function to regularize the sparsity of the solution. 
The larger the value of $\gamma_1$, the more sparse the final solution.
We use the minimization software NESTA for spectrum recovery \cite{NESTA}. 
To find the optimum value of $\gamma_1$, we first perform a cross-validation analysis to obtain an estimate.
Then, we scan $\gamma_1$ around this estimate and calculate the spectral reconstruction error $\epsilon \equiv ||S'-S||_2/||S||_2$, where $S$ is the input spectrum, and $S'$ is the  reconstructed spectrum. 
The minimum of $\epsilon$ sets the optimum value of $\gamma_1$ [see Fig. \ref{fig2}(c)]. 
Figure \ref{fig2}(a) shows a reconstructed spectrum that contains 50 narrow peaks over 100 nm range. 
The reconstruction for the entire spectrum takes less than 1 minute.
The Gaussian-shaped lines have unequal spacing, varying height and width. 
The expanded spectrum in Fig. \ref{fig2}(b) reveals that not only the position and height but also the spectral shape of each peak are accurately reconstructed.
Figure \ref{fig2}(d) shows that two lines separated by 0.03 nm are resolved. 

\begin{figure}[htbp]
	\centering
	\includegraphics[width=\linewidth]{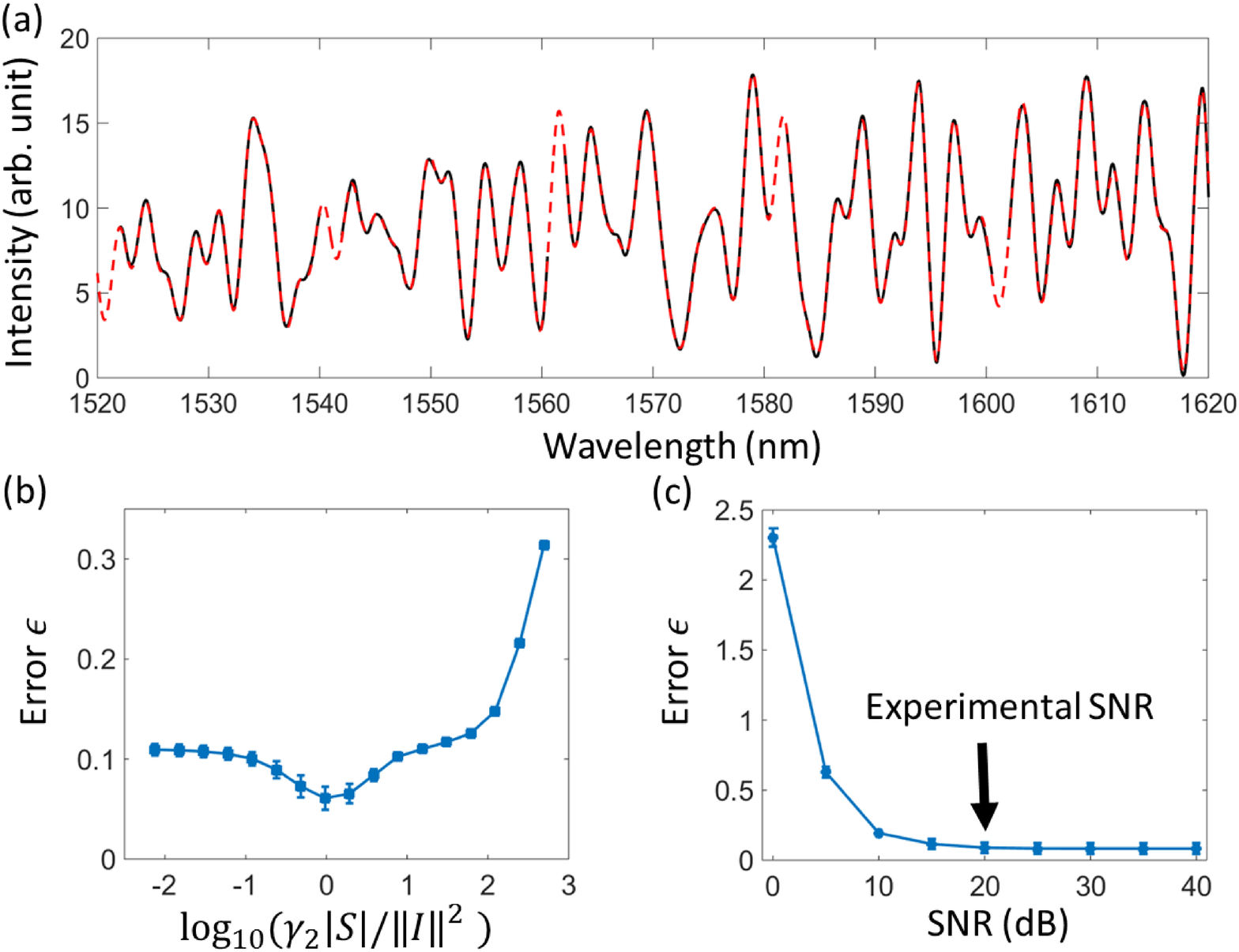}
	\caption{(Color online) (a) A reconstructed spectrum for a continuous broadband signal over 100nm. The black-solid line is the reconstructed spectrum $S'$ and the red-dashed line is the input spectrum $S$. The reconstruction error is $\epsilon = 0.05$. (b) Reconstruction error $\epsilon$ as a function of $\gamma_2$. (c) Reconstruction error $\epsilon$ as a function of the SNR of the probe speckle pattern. }
	\label{fig3}
\end{figure}

In a sparse spectrum, only a small number of spectral channels carry signals, the rest have vanishing amplitude, thus the speckle contrast is high and the optimization algorithm can quickly find the correct combination of spectral channels.  
However, a dense spectrum, e.g., a smooth, broadband spectrum, carries signals in almost all channels, and the speckle contrast is greatly reduced due to summation of many uncorrelated speckle patterns. 
The reconstructed spectrum obtained only from minimization of $||I-T \cdot S'||_2^2$ is often accompanied by rapid fluctuations, because there are too many spectral channels to optimize and the measurement noise is amplified. 

\indent To make the optimization efficient, we must reduce the number of components that need to be optimized. 
This can be done by converting the spectrum to a different basis, in which it becomes \textit{sparse}. 
For example, a discrete cosine transform (DCT) of a smooth, broadband spectrum gives a limited number of low-frequency components.
To reconstruct the sparse spectrum in the new basis, we minimize $ ||I- \tilde{T} \cdot \tilde{S'}||_2^2 + \gamma_2 ||\tilde{S'}||_1$, where $\tilde{S'} = D \cdot S'$, $\tilde{T} = T \cdot D^\mathsf{T}$, and $D$ is the discrete cosine transform matrix. 
The second term regularizes the sparsity of the solution. 
Figure \ref{fig3}(a) shows a reconstructed broad-band spectrum that spreads over 100 nm, the reconstruction error is $\epsilon$ = 0.05. 
The optimum value of $\gamma_2$ is chosen to minimize the reconstruction error [Fig. \ref{fig3}(b)]. 
The WDM used here is designed to avoid spectral overlap of different output channels, thus a ~ 2 nm spectral gap is left between neighboring channels. The spectral information within such gaps could not be collected, leading to missing data points in Fig. \ref{fig3}(a). 
This problem can be easily overcome by using a WDM with spectrally overlapping output channels. 

\begin{figure}[htbp!]
	\centering
	\includegraphics[width=\linewidth]{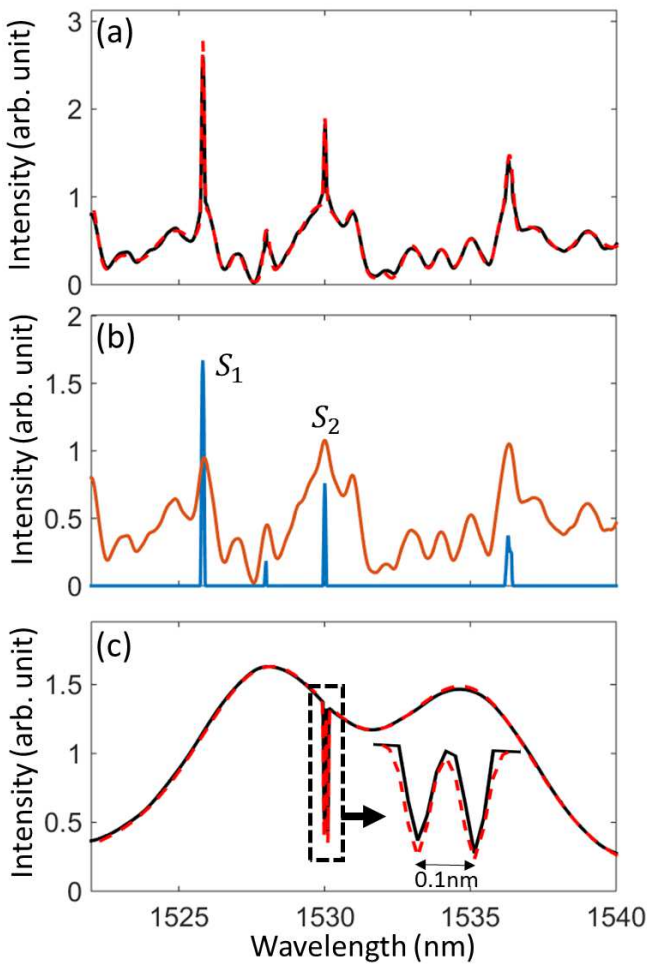}
	\caption{(Color online) (a) The reconstructed spectrum (black-solid) agrees well with the input spectrum that contains four narrow lines on top of a smooth varying signal (red-dashed). $\epsilon = 0.06$. (b) $S'_1$ is the reconstructed spectrum that is sparse in the wavelength domain and $S'_2$ is sparse in the DCT domain. (c) Reconstruction of a spectrum that contains two sharp dips separated by 0.1 nm on top of a broad background. $\epsilon = 0.003$. }
	\label{fig4}
\end{figure}

Next we investigate how the spectrum reconstruction is affected by the signal-to-noise ratio (SNR).  
To this end, we use the same transfer matrix for synthesis of the probe spectrum and for reconstruction, but we add white noise to the synthesized speckle pattern to simulate measurement noise. 
The SNR is given by the ratio of the integrated probe intensity (without noise) to the total noise intensity.  
Figure \ref{fig3}(c) plots the reconstruction error $\epsilon$ as a function of the SNR.
$\epsilon$ drops quickly as the SNR increases from 1dB to 10dB, then saturates to 0.05 beyond 10dB. 
Experimentally the SNR is 20dB, thus the spectrum can be accurately recovered. 

\indent We have shown that both narrow-band spectrum and broadband spectra can be accurately measured by the MMF spectrometer. 
The former bears similarity to a Raman spectrum, and the latter to a fluorescence spectrum. 
In addition, the probe spectrum may be a combination of both, e.g., a laser spectrum may contain sharp lasing lines on top of a smooth amplified spontaneous emission band. 
To reconstruct such spectrum, we will divide it to two parts, $S' = S'_1 + S'_2$,  $S'_1$ contains only sharp spectral features, and $S'_2$ the smooth varying spectrum. 
In other words, $S'_1$ is sparse in the wavelength domain and $S'_2$ is sparse in the DCT domain.
Hence, we minimize $ ||I-(T \cdot S'_1+ \tilde{T} \cdot \tilde{S'}_2)||_2^2 + \gamma_1 ||S'_1||_1 + \gamma_2 ||\tilde{S'}_2||_1$. 
Figure \ref{fig4}(a) shows a probe spectrum that contains four narrow lines on top of a smooth varying band. 
The reconstructed spectrum matches very well with the original one with an error $\epsilon = 0.06$. 
The two parts are plotted separately in Fig. \ref{fig4}(b), the reconstructed $S'_1$ has only four narrow lines, and $S'_2$ the smooth-varying spectrum without high-frequency noise.
Here we show the reconstructed spectrum in a single spectral window (measured from the speckle pattern of a single MMF), as each spectral window is effectively reconstructed individually. 
The same method can be applied to all other spectral windows.

In addition, we test another type of spectrum that has narrow dips on a broad background. 
It resembles the absorption spectrum taken with a broadband signal. 
As shown in Fig. \ref{fig4}(c), both the smooth background and the two closely-spaced sharp dips are accurately reconstructed, even though the wavelength spacing between the two dips is only 0.1 nm. 
The good agreement is confirmed by the small reconstruction error $\epsilon = 0.003$.  
These results confirm that the optimization algorithm can capture sharp spectral features while suppressing high-frequency noise in the smooth spectrum. 

In conclusion, by integrating a commercially available wavelength division multiplexer with multimode optical fibers, we have realized an all-fiber spectrometer with very broad bandwidth and fine spectral resolution.  
An efficient algorithm has been developed for accurate reconstruction of both sparse and dense spectra, or the combination of both. 
Since only five of the eight output ports of the WDM were used in this work, a further increase of bandwidth can be easily achieved. 
We note that the spectra covered by individual ports are reconstructed separately and in parallel, thus greatly reducing the complexity and enhancing the speed of reconstruction.
Moreover, the spectral resolution may be enhanced by using longer MMFs.  
Such a high-performance low-cost spectrometer will advance the development of portable spectroscopy systems for sensing and imaging applications. 

\section*{Funding Information}
This work is supported by the National Science Foundation (NSF) under the grant No. ECCS-1509361. 
\section*{Acknowledgement}




\end{document}